\documentclass{article}
\usepackage[preprint]{spconf}
\usepackage{amsmath,graphicx}
\usepackage{booktabs}
\usepackage{color, colortbl}
\usepackage{enumitem}
\usepackage[numbers,sort&compress]{natbib}

\newcolumntype{L}{>{\centering\arraybackslash}m{3cm}}

\newcommand{\method}{\textit{RespireNet}}

\title{RespireNet: A Deep Neural Network for Accurately Detecting Abnormal Lung Sounds in Limited Data Setting
}

\makeatletter
\newcommand{\printfnsymbol}[1]{%
  \textsuperscript{\@fnsymbol{#1}}%
}
\makeatother

\name{Siddhartha Gairola$^1$, Francis Tom$^2$, Nipun Kwatra$^1$ and Mohit Jain$^1$}
\address{Microsoft Research India$^1$, Microsoft$^2$}

\begin{document}

\maketitle
\begin{abstract}

Auscultation of respiratory sounds is the primary tool for screening and diagnosing lung diseases. Automated analysis, coupled with digital stethoscopes, can play a crucial role in enabling tele-screening of fatal lung diseases. Deep neural networks (DNNs) have shown a lot of promise for such problems, and are an obvious choice. However, DNNs are extremely data hungry, and the largest respiratory dataset ICBHI \citep{icbhi_17} has only 6898 breathing cycles, which is still small for training a satisfactory DNN model. In this work, \method{}, we propose a simple CNN-based model, along with a suite of novel techniques---device specific fine-tuning, concatenation-based augmentation, blank region clipping, and smart padding---enabling us to efficiently use the small-sized dataset. 
We perform extensive evaluation on the ICBHI dataset, and improve upon the state-of-the-art results for 4-class classification by 2.2\%.

\end{abstract}
\vspace{-2mm}
\section{Introduction}
\label{sec:introduction}
Respiratory diseases like asthma, chronic obstructive pulmonary disease (COPD), lower respiratory tract infection, lung cancer, and tuberculosis are the leading causes of death worldwide~\cite{who_17}, constituting four of the 12 most common causes of death.
Early diagnosis has been found to be crucial in limiting the spread of respiratory diseases, and their adverse effects on the length and quality of life.
Listening to chest sounds using a stethoscope is a standard method for screening and diagnosing lung diseases.
It provides a low cost and non-invasive screening methodology, avoiding the exposure risks of radiography and patient-compliance requirements associated with tests such as Spirometry. 

There are a few drawbacks of stethoscope-based diagnosis:
requirement of a trained medical professional to interpret auscultation signals, and subjectivity in interpretations causing  inter-listener variability.
These limitations are exacerbated in impoverished settings and during pandemic situations (such as COVID-19), due to shortage of expert medical professionals.
Automated analysis of respiratory sounds can alleviate these drawbacks, and also help in enabling tele-medicine applications to monitor patients outside a clinic by less-skilled workforce such as community health workers.

\begin{figure}[t]
\begin{center}
    \centering
    \includegraphics[width=1\linewidth]{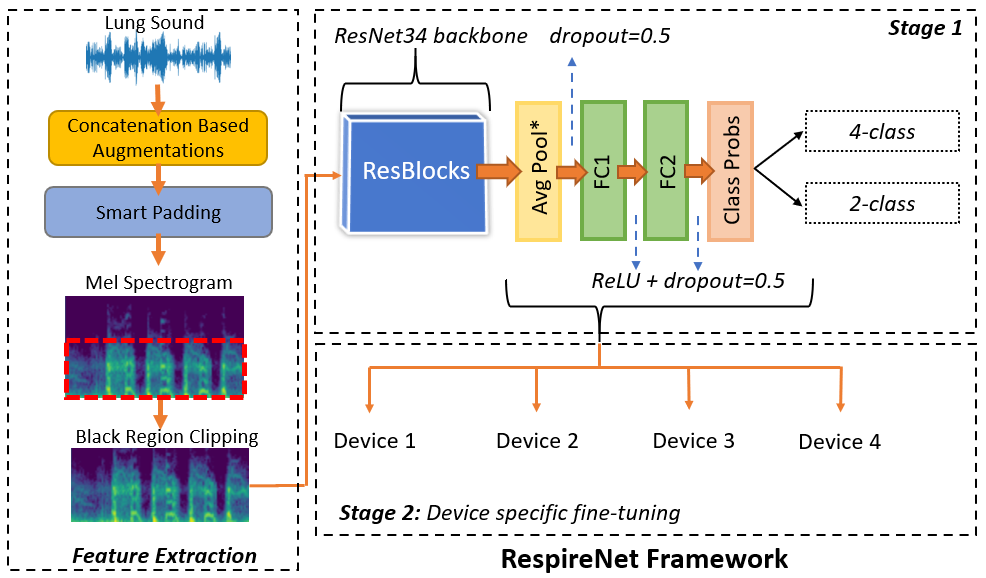}
\end{center}
\vspace{-5mm}
    \caption{\textit{Overview of proposed \method{} framework. We pre-process the audio signal (bandpass filtering, downsampling, normalization, etc.), apply concatenation-based augmentation and smart padding, and generate the mel-spectrogram. Blank region clipping is applied to remove blank regions in the high frequency ranges. The processed spectrogram is then used to train our DNN model via a two-stage training. Stage-1: the model is trained using entire train set. Stage-2: device specific fine-tuning which trains using subset of data corresponding to each device.}} 
    \label{fig:framework}
\vspace{-5mm}
\end{figure}

Algorithmic detection of lung diseases from respiratory sounds has been an active area of research \citep{polat_04, reichert_08}
especially with the advent of digital stethoscopes. Most of these works focus on detecting abnormal respiratory sounds of \textit{wheeze} and \textit{crackle}. Wheeze is a typical symptom of asthma and COPD. It is characterized by a high-pitched continuous sound in the frequency range of 100-2500Hz and duration above 80 msec \citep{lungausc_14, analysis_08}. Crackles, which are associated with COPD, chronic bronchitis, pneumonia and lung fibrosis \citep{analysis_11, automaticar_17}, have a discontinuous, non-tonal sound, around frequency of $\sim$650 Hz and duration of 5 msec (for fine crackles), or frequency of 100-500 Hz and duration of 15 msec (for coarse crackles).

Although early works focused on hand-crafted features and traditional machine learning~\cite{jako_18, chambres_18}, more recently, deep learning based methods have received the most attention~\cite{noise_rnn_koch_18,acharya_20, lungrn_20}.
For training DNNs, a time-frequency representation of the audio signal such as Mel-spectrograms \cite{shi_vggish_19, liu_detection_19, acharya_20}, stacked MFCC features \cite{shi_vggish_19, murat_17, perna_cnn_18, messner_18, noise_rnn_koch_18} or optimized S-transform spectrogram~\cite{triple_class_chen_19} is used. This 2D ``image" is then fed into CNNs \cite{perna_cnn_18, murat_17}, RNNs \cite{deepap_perna_19, noise_rnn_koch_18}, or hybrid CNN-RNNs \cite{acharya_20} to learn robust high dimensional representations.

It is well known that DNNs are data hungry and typically require large datasets to achieve good performance. In this work, we use the ICBHI challenge dataset \cite{icbhi_17}, a popular respiratory sound dataset.
In spite of being the largest publicly available dataset, it has only 6898 breathing cycle samples, which is quite small for training deep networks. Thus, a big focus of our work has been on developing a suite of techniques to help train DNNs in a data efficient manner. We found that a simple CNN architecture, such as ResNet, is adequate for achieving good accuracy. This is in contrast to prior work employing complex architectures like hybrid CNN-RNN \citep{acharya_20}, non-local block additions to CNNs \citep{lungrn_20}, etc.

In order to efficiently use the available data, we did extensive analysis of the ICBHI dataset. We found several characteristics of the data that might inhibit training DNNs effectively. For example, the dataset contains audio recordings from four different devices, with skewed distribution of samples across the devices, which makes it difficult for DNNs to generalize well across devices.
Similarly, the dataset has a skewed distribution across normal and abnormal classes, and varying lengths of audio samples.
We propose multiple novel techniques to address these problems---device specific fine-tuning, concatenation-based augmentation, blank region clipping, and smart padding. We perform extensive evaluation and ablation analysis of these techniques.

\noindent The main contributions of our work are:
\begin{enumerate}[nolistsep]
\item demonstration that a simple network architecture is sufficient for respiratory sound classification, and more focus is needed on making efficient use of available data.
\item a detailed analysis of the ICBHI dataset pointing out its characteristics impacting DNN training significantly.
\item a suite of techniques---device specific fine-tuning, concatenation-based augmentation, blank region clipping and smart padding---enabling efficient dataset usage. These techniques are orthogonal to the choice of network architecture and should be easy to incorporate in other networks. 
\end{enumerate}
\vspace{-2mm}
\section{Method}
\label{sec:method}
\vspace{-2mm}
\noindent\textit{Dataset:} We perform all evaluations on the ICBHI scientific challenge respiratory sound dataset \cite{icbhi_17, icbhi_19}. It is one of the largest publicly available respiratory datasets. The dataset comprises of 920 recordings from 126 patients with a combined total duration of 5.5 hours. Each breathing cycle in a recording is annotated by an expert as one of the four classes: \textit{normal}, \textit{crackle}, \textit{wheeze}, or \textit{both} (crackle and wheeze). The dataset comprises of recordings from four different devices\footnote{The four devices used for recordings are AKGC417L Microphone, 3M Littmann Classic II SE Stethoscope, 3M Litmmann 3200 Electronic Stethoscope, and WelchAllyn Meditron Master Elite Electronic Stethoscope} from hospitals in Portugal and Greece. For every patient, data was recorded at seven different body locations.

\smallskip\noindent\textit{Pre-processing:} The sampling rate of recordings in the dataset varies from 4 kHz to 44.1 kHz. To standardize, we down-sample the recordings to 4 kHz and apply a 5-th order Butterworth band-pass filter to remove noise (heartbeat, background speech, etc.). We also apply standard normalization on the input signal to map the values within the range (-1, 1). The audio signal is then converted into a Mel-spectrogram, which is fed into our DNN.

\smallskip\noindent\textit{Network architecture:} We use a CNN-based network, \textit{ResNet-34}, followed by two \textit{128-d} fully connected linear layers with \textit{ReLU} activations. The last layer applies \textit{softmax activation} to model classwise probabilities. Dropout is added to the fully-connected layers to prevent overfitting. The network is trained via a standard categorical cross-entropy loss to minimize the loss for multi-class classification. The overall framework and architecture is illustrated in Figure~\ref{fig:framework}.

\vspace{-2mm}
\subsection{Efficient Dataset Utilization}
\vspace{-2mm}
Even though ICBHI is the largest publicly available dataset with 6898 samples, it is still relatively small for training DNNs effectively. Thus, a major focus of our work has been to develop techniques to efficiently use the available samples. We extensively analyzed the dataset to identify dataset characteristics that inhibit training DNNs effectively, and propose solutions to overcome the same.

The first commonly used technique we apply is \textit{transfer learning}, where we initialize our network with weights of a pre-trained \textit{ResNet-34} network on ImageNet \cite{ILSVRC15}. This is followed by our training where we train the entire network end-to-end. Interestingly, even though ImageNet dataset is very different from the spectrograms which our network sees, we still found this initialization to help significantly. Most likely, low level features such as edge-detection are still similar and thus ``transfer'' well.

\begin{figure}[!htbp]
\begin{center}
    \centering
    \includegraphics[width=\linewidth]{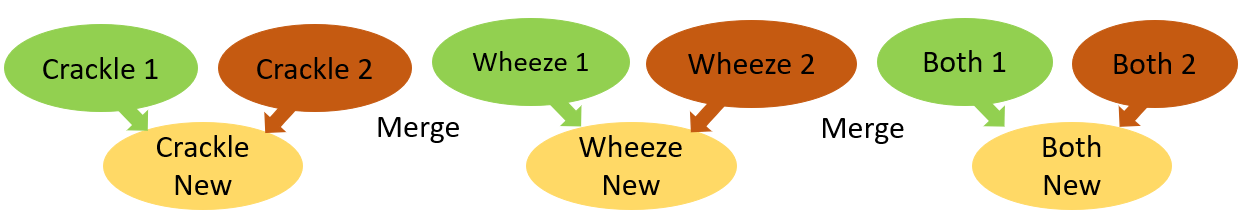}
\end{center}
\vspace{-5mm}
    \caption{\textit{Proposed concatenation-based augmentation.}} 
    \label{fig:new_aug_scheme}
\end{figure}

\smallskip\noindent\textbf{Concatenation-based Augmentation:} Like most medical datasets, ICBHI dataset has a huge class imbalance, with the \textit{normal} class accounting for 53\% of the samples. To prevent the model from overfitting on abnormal classes, we experimented with several data augmentation techniques. We first apply standard audio augmentation techniques, such as noise addition, speed variation, random shifting, pitch shift, etc., and also use a weighted random sampler to sample mini-batches uniformly from each class. These standard techniques help a little, but to further improve generalization of the underrepresented classes (\textit{wheeze}, \textit{crackle}, \textit{both}), we developed a concatenation based augmentation technique where we generate a new sample of a class by randomly sampling two samples of the same class and concatenating them (see Figure~\ref{fig:new_aug_scheme}). This scheme led to a non-trivial improvement in the classification accuracy of abnormal classes.

\noindent

\smallskip\noindent\textbf{Smart Padding:}
The breathing cycle length varies across patients as well as within a patient depending on various factors (\textit{e.g.}, breathing rate can increase moderately during fever).
In the ICBHI dataset, the length of breathing cycles ranges from 0.2s to 16.2s with a mean cycle length of 2.7s.
This poses a problem while training our network as it expects a fixed size input\footnote{CNNs can be made size agnostic by using adaptive average pooling, but that typically hurts accuracy.}. The standard way to handle this is to pad the audio signal to a fixed size via \textit{zero-padding} or \textit{reflection} based padding. We propose a novel \textit{smart padding} scheme, which uses a variant of the \textit{augmentation} scheme described above.
For each data sample, \textit{smart padding} first looks at the breathing cycle sample for the same patient taken just before and after the current one. If this neighbouring cycle is of the same class or of the \textit{normal} class, we concatenate the current sample with it. If not, we pad by copying the same cycle again. We continue this process until we reach our desired size. This \textit{smart padding} scheme also augments the data and helps prevent overfitting. We experimented with different input lengths, and found a 7s window to perform best. A small window led to clipping of samples, thus loosing valuable information in an already scarce dataset, while a very large window caused repetition leading to degraded performance.

\begin{figure}[!hbtp]
\begin{center}
    \centering
    \includegraphics[width=\linewidth]{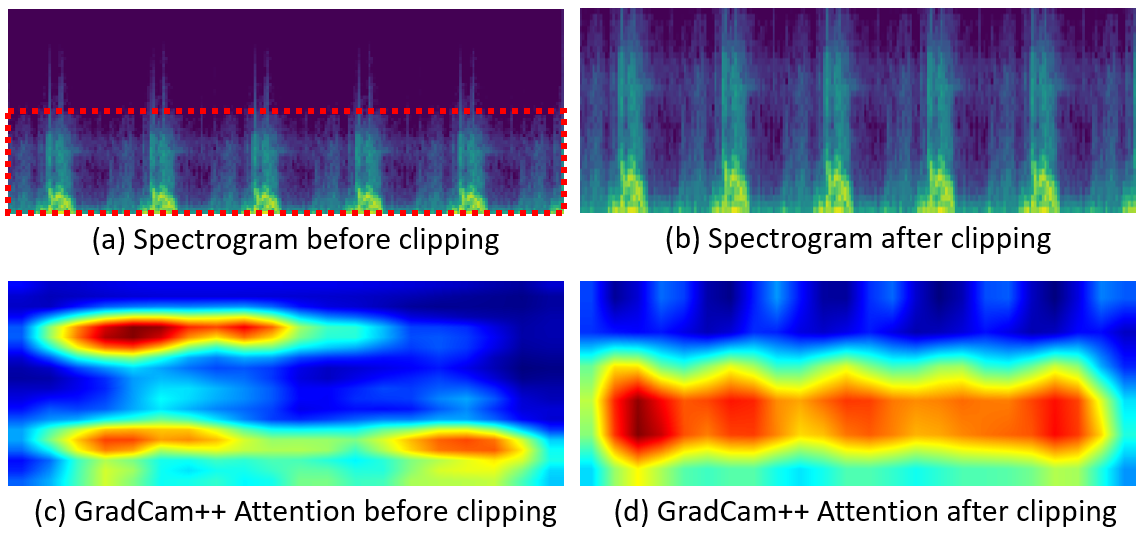}
\end{center}
\vspace{-5mm}
    \caption{\textit{Blank region clipping: The network attention~\cite{gradcampp} starts focusing more on the bottom half of the spectrogram, instead of blank spaces after clipping.}}
    \label{fig:black_clip}
\end{figure}

\noindent\textbf{Blank Region Clipping:}
On analyzing samples using Grad-Cam++~\cite{gradcampp} which our base model mis-classified, we found notable black regions\footnote{Black region in a spectrogram means that the audio signal has zero energy in the corresponding audio frequency range.} at higher frequency regions of their spectrograms (Figure~\ref{fig:black_clip}). On further analysis, we found that many samples, and in particular 100\% of the Litt3200 device samples, had blank region in the 1500-2000Hz frequency range. Since this was adversely affecting our network performance, we selectively clip off the blank rows from the high frequency regions of such spectrograms. This ensures that the network focuses on the region of interest leading to improved performance. Figure~\ref{fig:black_clip} shows this in action.

\smallskip\noindent\textbf{Device Specific Fine-tuning:} The ICBHI dataset has samples from 4 different devices. We found that the distribution of samples across devices is heavily skewed, \textit{e.g.} the AKGC417L Microphone alone contributes to 63\% of the samples. Since each device has different audio characteristics, the DNN may fail to generalize across devices, especially for the underrepresented devices in the already small dataset. To verify this, we divided the test set into 4 subsets depending on their device type, and compute the accuracy of abnormal class samples in each subset. As expected, we found the classification accuracy to be strongly correlated with the training set size of the corresponding device.
To address this, we first train a common model with the full training data (stage-1, Figure~\ref{fig:framework}). We then make 4 copies of this model and \textit{fine-tune} (stage-2) them for each device separately by using only the subset of training data for that device. We found this approach to significantly improve the performance, especially for the underrepresented devices.
\vspace{-2mm}
\section{Experiments}
\vspace{-2mm}

We evaluate the performance of our framework on the respiratory anomaly classification task proposed in the ICBHI challenge~\cite{icbhi_17}. This is further divided into two subtasks: (i) classify a breathing cycle into one of the four classes--\textit{normal(N), crackle(C), wheeze(W), both(B)}, and (ii) classify a breathing cycle into \textit{normal or anomalous} class, where $anomalous = \{crackle, wheeze, both\}$.
Our evaluation method is same as the one proposed in the original ICBHI challenge. The final score is computed as the mean of Sensitivity $S_e=\frac{P_c + P_w + P_b}{N_c + N_w + N_b}$ and Specificity $S_p=\frac{P_n}{N_n}$, where $P_i$ and $N_i$ are the number of correctly classified and total number of samples in class $i$, respectively ($i \in \{normal, crackle, wheeze, both\}$).
For the 2-class case, we adopt the anomalous and normal class scores as $S_e$ and $S_p$ respectively, and the score is computed as their mean.

We compare our performance using the above evaluation metric on two dataset divisions: the official 60-40$\%$ split \cite{icbhi_17} and the 80-20$\%$ split \cite{lungbrn_19, lungrn_20, acharya_20} for train-test\footnote{For both the splits, the train and test set are patient-wise disjoint.}. The Sensitivity $S_e$, Specificity $S_p$ and ICBHI Score values are reported in Table \ref{tab:result_comparison}. \method{} achieves state-of-the-art (SOTA) in both train-test split divisions, and outperforms SOTA \cite{lungrn_20} on the official split (60-40) by $4\%$ and SOTA \cite{acharya_20} on the 80-20 split by $2.2\%$. Further, \method{} achieves a score of $77\%$ on the 2-class classification task, achieving the new SOTA.

\smallskip\noindent\textit{Implementation Details:} We train our models on a Tesla v100 GPU on a Microsoft Azure VM. We used the SGD optimizer with momentum of 0.9, and a batch size of 64. We used a fixed learning rate of 1e-3 for stage-1 and 1e-4 for stage-2 of training. Stage-1 was trained for 200 epochs. The highest validation checkpoint from stage-1 was used to train stage-2 for another 50 epochs for each device.

\begin{table}[t]
\centering
 \resizebox{\linewidth}{!}{
\begin{tabular}{ c | c | c c c}
\toprule
 \textbf{Split \& Task}& \textbf{Method} & \boldmath$S_p$ & \boldmath$S_e$ & \textbf{Score}\\
\cmidrule(lr){1-1}
\cmidrule(lr){2-2}
\cmidrule(lr){3-5}

\textbf{60$/$40 Split} & Jakovljevic et al. \cite{jako_18} &  - & - & 39.5\% \\
\textbf{\&} & \citet{chambres_18}  & 78.1\% & 20.8\% & 49.4\% \\
\textbf{4-class} & \citet{serbes_18} & - & - & 49.9\% \\
& \citet{lungbrn_19}  & 69.2\% & 31.1\% & 50.2\% \\
& \citet{lungrn_20}  & 63.2\% & 41.3\% & 52.3\% \\
& CNN (ours) & 71.4\% & 39.0\% & 55.2\% \\
& CNN+CBA+BRC (ours)  & 71.8\% & 39.6\% & 55.7\% \\
& CNN+CBA+BRC+FT (ours) & 72.3\% & 40.1\% & \textbf{56.2}\% \\

\cmidrule(lr){1-1}
\cmidrule(lr){2-2}
\cmidrule(lr){3-5}

\textbf{80$/$20 Split} & \citet{noise_rnn_koch_18} & 73.0\% & 58.4\% & 65.7 \%\\
\textbf{\&} & Acharya et al. \cite{acharya_20}  &  84.1\% & 48.6\% & 66.3\% \\
\textbf{4-class} & \citet{lungrn_20} & 64.7\% & 63.7\% & 64.2\% \\
& CNN (ours) & 78.8\% & 53.6\% & 66.2\% \\
& CNN+CBA+BRC (ours)  & 79.7\% & 54.4\% & 67.1\% \\
& CNN+CBA+BRC+FT (ours) & 83.3\% & 53.7\% & \textbf{68.5}\% \\

\cmidrule(lr){1-1}
\cmidrule(lr){2-2}
\cmidrule(lr){3-5}

\textbf{80$/$20 Split} & CNN (ours) & 83.3\% & 60.5\% & 71.9\% \\
\textbf{\&} & CNN+CBA+BRC (ours) & 76.4\% & 71.0\% & 73.7\% \\
\textbf{2-class} & CNN+CBA+BRC+FT (ours)  & 80.9\% & 73.1\% & \textbf{77.0\%} \\

 \bottomrule
\end{tabular}
 }
\vspace{-2mm}
\caption{\textit{Performance comparison of the proposed model with the state-of-the-art systems following random splits. We see significant improvements from our proposed techniques: concatenation-based augmentation (CBA), blank region clipping (BRC) and device specific fine-tuning (FT).}}
\label{tab:result_comparison}
\end{table}

We further analyze the effect of our novel proposed techniques by conducting an ablation analysis on the 4-class classification task on the 80/20 split.

\smallskip\noindent\textit{Concatenation-based Augmentation:}
Due to the small size of abnormal samples in the dataset, our model tends to overfit on the abnormal classes quickly, and achieved a score of $62.2\%$. Standard augmentations (noise addition, etc.) improved the score to $66.2\%$, which further improved to $66.8\%$ with our concatenation-based augmentation. Also, most of the gain came from improved accuracy of the abnormal classes, where the sensitivity increased by 1.5\%. This demonstrates that our augmentation scheme to generate novel samples for the abnormal class helps the model generalize better.

\begin{table}[t]
\centering
 \resizebox{\linewidth}{!}{
\begin{tabular}{ c | c c c c c c c c c}
\toprule
\textbf{Length.} & \textbf{1 sec} & \textbf{2 sec} & \textbf{3 sec} & \textbf{4 sec} & \textbf{5 sec} & \textbf{6 sec} & \textbf{7 sec} & \textbf{8 sec} & \textbf{9 sec}\\ 
\cmidrule(lr){1-1}
\cmidrule(lr){2-10}

\textbf{Scores} & 56.6 & 59.0 & 60.3 & 61.1 & 62.3 & 64.4 & \textbf{66.2} & 65.1 & 65.5\\

 \bottomrule
\end{tabular}
}
\vspace{-2mm}
\caption{\textit{Input length size  vs classification score.}}
\vspace*{-12pt}
\label{tab:time-slice_ablation}
\end{table}
\smallskip\noindent\textit{Smart Padding:} The length of breathing cycle in the dataset has a wide variation, thus we need to pad the shorter samples and clip the longer ones to match the input length of the network. We experimented with different input lengths and found that a 7s length performed optimally (see Table~\ref{tab:time-slice_ablation}). Since the average cycle length is 2.7s, padding became crucial as a majority of the inputs need padding. We found the padding scheme to have a significant impact on accuracy. For the base model, \textit{smart padding} improves accuracy over \textit{zero-padding} and \textit{reflection-based} padding by 5\% and 2\% respectively. This demonstrates the effectiveness of our padding scheme, which incorporates data augmentation for padding, rather than plain copying or adding zeros.

\smallskip\noindent\textit{Blank Region Clipping:} This provided an improvement of $0.5\%$ over the base model score of $66.2\%$. When combined with our proposed augmentation, it helped achieve a score of $67.1\%$, outperforming the current SOTA~\cite{acharya_20} by $0.8\%$.

\smallskip\noindent\textit{Device specific fine-tuning:} We found that the large skew in sample distribution across devices caused the model to not generalize well for under-represented devices. Our device specific fine-tuning scheme helped significantly, resulting in an improvement of $1.4\%$ in the final ICBHI score. We also observed that this fine-tuning disproportionally helped the under-represented classes. Table~\ref{tab:scores_device_wise} shows that devices with fewer samples had $\sim$9\% increase in their scores.

\begin{table}[htbp]
    \centering
    \small
    \begin{tabular}{c | c | c}
    \toprule
	\textbf{Device} & \textbf{\% Samples}  & \textbf{Score Improvement} \\
    \cmidrule(lr){1-1}
    \cmidrule(lr){2-2}
    \cmidrule(lr){3-3}
    AKGC417L & 63\% & 1.7\%\\

    Meditron & 21\% &  1.6\%\\
   
    Litt3200 & 9\% &  9.3\%\\

    LittC2SE & 7\% & 8.6\%\\
    \bottomrule
    \end{tabular}
    \vspace{-2mm}
    \caption{\textit{Device specific fine-tuning: The devices with small number of samples show a big improvelment in their scores.}}
    \label{tab:scores_device_wise}
    \vspace{-12pt}
\end{table}

\vspace{-3mm}
\section{Related Work}
\vspace{-2mm}
Recently, there has been a lot of interest in using deep learning models for respiratory sounds classification~\cite{acharya_20, lungrn_20, noise_rnn_koch_18}. It has outperformed statistical methods (HMM-GMM) \cite{jako_18} and traditional machine learning methods (boosted decision trees, SVM) \cite{chambres_18, serbes_18}. 
In these deep learning based approaches, a time-frequency representation of the audio signal is provided as input to the model. \citet{noise_rnn_koch_18} propose a deep recurrent network with a noise masking intermediate step for the four class classification task, obtaining a score of $65.7\%$ on the 80-20 split. However the paper omits the details regarding noise label generation~\cite{acharya_20},
thus making it hard to reproduce.
Deep residual networks and optimized S-transform based features are used by \citet{triple_class_chen_19} for three-class classification of anomalies in lung sounds. The model is trained and tested on a smaller subset of the ICBHI dataset on a 70-30 split and achieve a score of $98\%$.

\citet{acharya_20} propose a Mel-spectrogram based hybrid CNN-RNN model with patient-specific model tuning, achieving a score of $66.3\%$ on 4-class and 80-20 split. \citet{lungrn_20} introduce LungRN+NL which incorporates a non-local block in the ResNet architecture and apply mixup augmentations to address the data imbalance problem and improve the model's robustness, achieving sensitivity of $63.7\%$.
However, none of these approaches focus on characteristics of the ICBHI dataset, which we exploit to improve performance.

\vspace{-2mm}
\section{Conclusion and Future Work}
\vspace{-2mm}

The paper proposes \method{} a simple CNN-based model, along with a set of novel techniques---device specific fine-tuning, concatenation-based augmentation, blank region clipping, and smart padding---enabling us to effectively utilize a small-sized dataset for accurate abnormality detection in lung sounds. Our proposed method achieved a new SOTA for the ICBHI dataset, on both the 2-class and 4-class classification tasks. Further, our proposed techniques are orthogonal to the choice of network architecture and should be easy to incorporate within other frameworks.

The current performance limit of the 4-class classification task can be mainly attributed to the small size of the ICBHI dataset, and the variation among the recording devices. Furthermore, there is lack of standardization in the 80-20 split and we found variance in the results based on the particular split. In future, we would recommend that the community should focus on capturing a larger dataset, while taking care of the issues raised in this paper.

\clearpage

{\small
\bibliography{paper}
}

\clearpage
\section{Supplementary Material}
This supplementary material includes some other details about the dataset, and additional results which could not be accommodated in the main paper.

\subsection{Dataset Details}
\begin{figure}[!hbtp]
\begin{center}
    \centering
    \includegraphics[width=\linewidth]{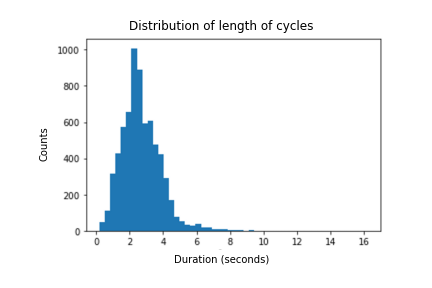}
\end{center}
\vspace{-8mm}
    \caption{\textit{Distribution of length of cycles across samples. 65\% of the samples have a cycle length of $<3$ seconds, and 33\% of the samples have a cyle length between 4-6 seconds.}} 
    \label{fig:dist_cycle_length}
\end{figure}

The 2017 ICBHI dataset \cite{icbhi_17} comprises of 920 recordings from 126 patients with a combined total duration of 5.5 hours. Each breathing cycle in a recording is annotated by a single expert as one of the four classes: \textit{normal, crackle, wheeze or both (crackle and wheeze)}. These cycles have various recording lengths (see Figure \ref{fig:dist_cycle_length}) ranging from 0.2s to 16.2s (mean cycle length is 2.7s) and the number of cycles is imbalanced across the four classes (i.e. 3642, 1864, 886, 506 cycles for \textit{normal, crackle, wheeze and both} classes respectively).

The dataset consists of sound recordings from four devices \textit{AKGC417L Microphone, 3M Littmann Classic II SE Stethoscope, 3M Litmmann 3200 Electronic Stethoscope and WelchAllyn Meditron Master Elite Electronic Stethoscope} and is not balanced across patients as well as number of breathing cycles (see Tables \ref{tab:cyles_class_device}, \ref{tab:per_class_device}). This creates a skew in the data distribution and has an adverse impact on the performance of the model as discussed in the analysis earlier.
\begin{table}[!htbp]
    \centering
    \small
    \begin{tabular}{c | c | c c c c c}
    \toprule
	\textbf{Device} & \multicolumn{1}{m{1cm}|}{\textbf{Patient Count}*} & \textbf{N} & \textbf{C} & \textbf{W}	& \textbf{B} & \textbf{Total}\\
    \cmidrule(lr){1-1}
    \cmidrule(lr){2-2}
    \cmidrule(lr){3-7}
    AKGC417L & 32 & 1922 & 1543 & 500 & 381 & 4346\\
    Meditron & 64 & 1037 & 215 & 148 & 56 & 1456\\	
    Litt3200 & 11 & 347 & 77 & 126 & 44 & 594\\
    LittC2SE & 23 & 336 & 29 & 112 & 25 & 502\\
    \bottomrule
    \end{tabular}
    \caption{\textit{Number of breathing cycles across classes and devices, along with the distribution of patients across devices.
    \newline
    *Number of patients total to 130 instead of 126 as some of the devices have an overalap with the patients.}
    }
    \label{tab:cyles_class_device}
\end{table}

\begin{table}[!htbp]
    \centering
    \begin{tabular}{c | c c c c}
    \toprule
	\textbf{Device} & \textbf{N} & \textbf{C} & \textbf{W}	& \textbf{B}\\
    \cmidrule(lr){1-1}
    \cmidrule(lr){2-5}
	AKGC417L & 0.53 & 0.83 & 0.56 & 0.75\\
    Meditron & 0.28 & 0.11 & 0.17 & 0.11\\
	Litt3200 & 0.10 & 0.02 & 0.14 & 0.09\\
	LittC2SE & 0.09 & 0.04 & 0.13	& 0.05\\
    \bottomrule
    \end{tabular}
    \caption{\textit{Distribution of breathing cycles across classes and devices.}}
    \label{tab:per_class_device}
\end{table}

For creating the splits we perform sample 80-20 w.r.t number of patients. From the numbers in Table \ref{tab:cyles_class_device}, we have 64 patients from \textit{Meditron} device but only 1468 breathing cycles (22.9 breathing cycles per patient on an average), whereas for \textit{AKGC417L} device we have 32 patients and 4364 breathing cycles (136.4 breathing cycles per patient on an average). This depicts the huge skew in the splits across devices and patients. Further there is also a skew between abnormal classes across devices: The majority of \textit{crackle} class (83\% of the total samples) is found within the \textit{AKGC417L} device whereas \textit{wheeze} and \textit{both} have different proportions across devices. 

\subsection{Additional Results}

\smallskip \textit{Single Device Training} We train our model only on samples from the \textit{AKGC417L} device. Table \ref{tab:single_device_training} depicts the test performance on the 4 different devices. This demonstrates that the training only on a single device, does not translate well across the other devices, thus further motivating the use of \textit{device specific fine-tuning}.

\begin{table}[!htbp]
    \centering
    \small
    \resizebox{0.8\linewidth}{!}{
    \begin{tabular}{c | c c c c}
    \toprule
	\textbf{Device} & \textbf{Normal} & \textbf{Crackle} & \textbf{Wheeze}	& \textbf{Both}\\
    \cmidrule(lr){1-1}
    \cmidrule(lr){2-5}
    
    AKGC417L & 61.3\% & 77.5\% & 23.8\% & 28.2\%\\
    Meditron& 47.3\% & 69.2\% & 26.3\% & 0.0\%\\
	Litt3200 & 51.2\% & 66.7\% & 20.7\% & 66.7\%\\
	LittC2SE & 16.8\% & 22.2\% & 0.0\% & 0.0\%\\

    \bottomrule
    \end{tabular}
    }
    \vspace{-2mm}
    \caption{\textit{Scores device wise for each class when trained only on AKGC417L. Overall \textbf{Score:} 53.0\%, \textbf{Sensitivity:} 55.7\% and \textbf{Specificity:} 50.3\%.}}
    \label{tab:single_device_training}
\end{table}

\begin{figure*}[!hbtp]
\begin{center}
    \centering
    \includegraphics[width=\textwidth]{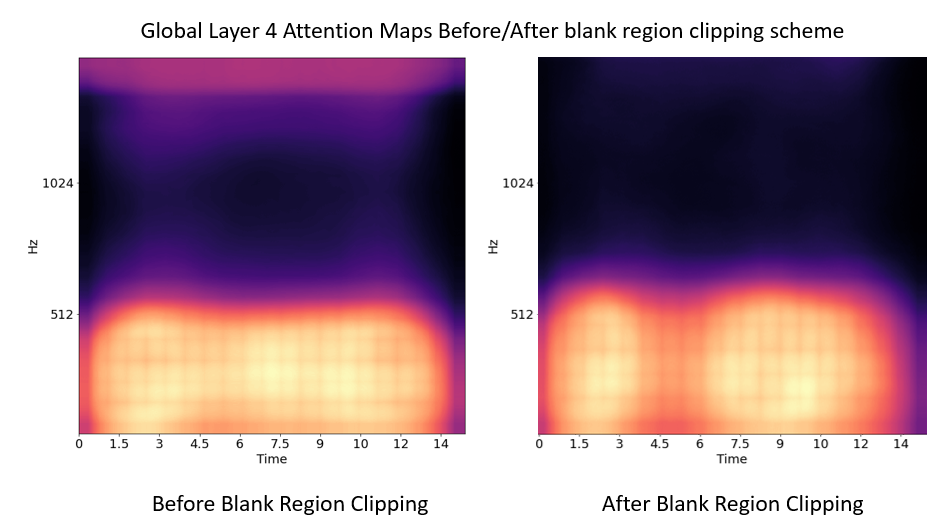}
\end{center}
\vspace{-4mm}
    \caption{\textit{Global average of attention maps computed using Grad-Cam++\cite{gradcampp} for samples in the test split before and after employing the blank region clipping scheme during network training.}} 
    \label{fig:attention_maps}
\end{figure*}

\smallskip \textit{Attention Map Visualization} Figure \ref{fig:attention_maps} depicts  global average of attention maps computed (for layer-4 of \textit{ResNet34}) using Grad-Cam++\cite{gradcampp} for 1370 samples in the test split before and after employing the \textit{blank region clipping} scheme during network training. It can be observed that the network starts focusing more on the bottom half of the spectrogram, instead of blank spaces after using blank region clipping. This demonstrates the efficacy of using the proposed \textit{blank region clipping} scheme which also results in improved performance.

\begin{figure*}[!hbtp]
\begin{center}
    \centering
    \includegraphics[width=\textwidth]{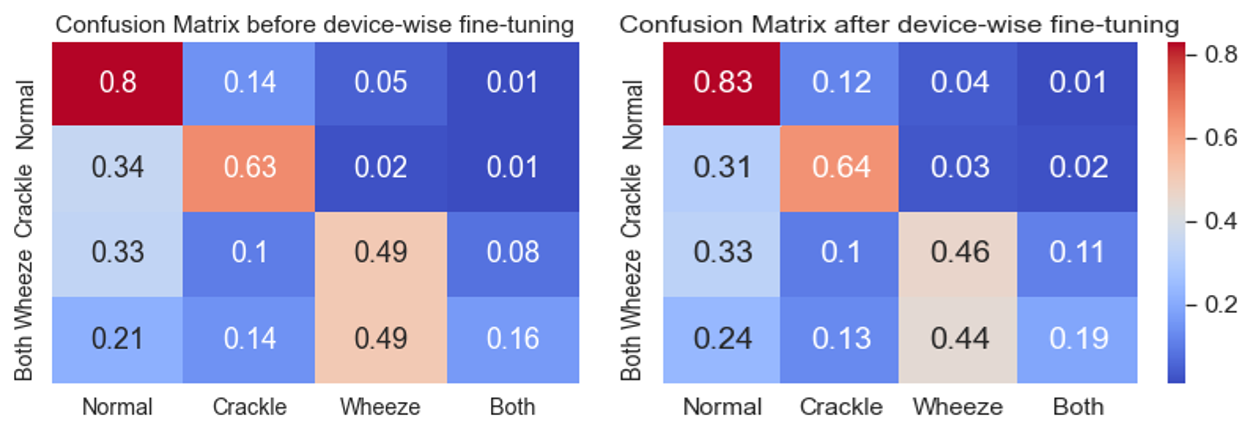}
\end{center}
\vspace{-2mm}
    \caption{\textit{Confusion matrices before and after device-wise fine-tuning.}} 
    \label{fig:confusion_matrix}
\end{figure*}

\smallskip \textit{Confusion Matrix} Figure \ref{fig:confusion_matrix} shows the confusion matrix before and after \textit{device specific fine-tuning}.

\end{document}